\documentclass[amsmath, amssymb, reprint, superscriptaddress, aip, apl]{revtex4-1}
\usepackage[utf8]{inputenc}
\usepackage{amsmath}
\usepackage{amsfonts}
\usepackage{amssymb}
\usepackage{lmodern}
\usepackage{graphicx}
\usepackage{tabularx}
\usepackage{booktabs}
\usepackage[separate-uncertainty = true,multi-part-units=single]{siunitx}
\usepackage{color}
\usepackage{nicefrac}
\usepackage{glossaries}
\usepackage{cleveref}

\AtBeginDocument{
    \heavyrulewidth=.08em
    \lightrulewidth=.05em
    \cmidrulewidth=.03em
    \belowrulesep=.65ex
    \belowbottomsep=0pt
    \aboverulesep=.4ex
    \abovetopsep=0pt
    \cmidrulesep=\doublerulesep
    \cmidrulekern=.5em
    \defaultaddspace=.5em
}

\newacronym[longplural={valence band offsets}]{vbo}{VBO}{valence band offset}
\newacronym[longplural={conduction band offsets}]{cbo}{CBO}{conduction band offset}
\newacronym{dft}{DFT}{density functional theory}
\newacronym[longplural={valence band maxima}]{vbm}{VBM}{valence band maximum}
\newacronym[longplural={conduction band maxima}]{cbm}{CBM}{conduction band maximum}
\newacronym{lda}{LDA}{local density approximation}
\newacronym{gga}{GGA}{generalized gradient approximation}
\newacronym{so}{SO}{spin-orbit}
\newacronym{wqw}{WQW}{W-quantum well}
\newacronym{sqs}{SQS}{special quasi-random structure}
\newacronym{si}{Si}{silicon}
\newacronym{cmos}{CMOS}{complementary metal-oxide semiconductor}
\newacronym{gap}{GaP}{gallium phosphide}
\newacronym{ganasp}{Ga(N,As,P)}{gallium phosphide arsenide nitride}
\newacronym[longplural={static atomic displacements}]{sad}{SAD}{static atomic displacement}
\newacronym{ic}{IC}{integrated circuit}
\newacronym{rt}{RT}{room temperature}
\newacronym{ple}{PLE}{photoluminescence excitation}
\newacronym{pl}{PL}{photoluminescence}
\newacronym{hh}{hh}{heavy hole}
\newacronym{lh}{lh}{light hole}
\newacronym{bac}{BAC}{band anti-crossing}
\newacronym{vca}{VCA}{virtual crystal approximation}
\newacronym{mqw}{MQW}{multi quantum well}
\newacronym{ldos}{LDOS}{localized density of states}

\begin{document}

\title{\emph{ab-initio} calculation of band alignments for opto-electronic simulations}

\author{Jan Oliver Oelerich}
\email{jan.oliver.oelerich@physik.uni-marburg.de}
\affiliation{Faculty of Physics and Materials Sciences Center, Philipps-Universität Marburg, Germany}

\author{Maria J. Weseloh}
\affiliation{Faculty of Physics and Materials Sciences Center, Philipps-Universität Marburg, Germany}

\author{Kerstin Volz}
\affiliation{Faculty of Physics and Materials Sciences Center, Philipps-Universität Marburg, Germany}

\author{Stephan W. Koch}
\affiliation{Faculty of Physics and Materials Sciences Center, Philipps-Universität Marburg, Germany}



\date{\today}

\begin{abstract}
    A modified core-to-valence band maximum approach is applied to calculate band offsets of strained III/V semiconductor hetero junctions. The method is used for the analysis of (In,Ga)As/GaAs/Ga(As,Sb) multi-quantum well structures. The obtained offsets and the resulting bandstructure are used as input for the microscopic calculation of photoluminescence spectra yielding very good agreement with recent experimental results. 
\end{abstract}

\maketitle

\section{Introduction}
\label{sec:intro}

Semiconductor hetero structures are the basic building blocks of many opto-electronic devices, such as solar cells, semiconductor sensors, or laser diodes. By choosing appropriate alloys and layer structures, device makers have great flexibility in engineering the optical and electronic properties to meet their requirements. However, the parameter space for designing such structures is too large for simple experimental trial-and-error. It is therefore necessary to thoroughly understand the physics of these devices, and to be able to predict their performance with accurate simulation methods. 

One fundamental requirement for the reliable prediction of the opto-electronic semiconductor hetero structure properties is a detailed knowledge of the electronic band structure throughout the device~\cite{doi:10.1142/7184}. Because energy bands are a property of the infinite system, complex layered hetero structures cannot be modeled directly using standard first principles approaches. Instead, one typically uses the so-called envelope function approximation where one keeps the bandstructure of the infinite bulk materials in the plane of the layers and accounts for the finite thickness in growth direction such that the hetero structure can be approximated by stacking these layers on top of each other~\cite{Winkler1993}. The infinite band structures of the reference bulk systems are then modified to take into account lattice strain imposed by the substrate, and quantum confinement arising from finite layer thicknesses. 

In a second step, the bands of adjacent layers are connected at the structure interfaces, leading to the devices' position-dependent overall band structure. The critical parameter required to connect the band structures at an interface is the relative energy offset of their \glspl{vbm}, the \gls{vbo}~\cite{FRANCIOSI1996}. It determines both, transport across the interface as well as quantum confinement of the layers, making it one of the most important parameters for the design of hetero structures. 

Estimations of the band alignment at semiconductor interfaces from first principles band structure calculations have a long history with various different suggested methods~\cite{Tersoff1984,Wei1998,VanDeWalle2006,Komsa2008}. The state of the art is to combine calculations of bulk-like properties of the constituents with information about the interface. This typically requires performing at least three separate calculations (bulk material $X$, bulk material $Y$, interface $XY$) but has the advantage of yielding the true bulk-like \gls{vbo} with manageable computational effort~\cite{Komsa2008}. 

In this paper, we use the core-to-\gls{vbm} method introduced by Wei and Zunger~\cite{Wei1998}, modified to take into account anisotropic strain in the grown layers as well as the exact chemical environment of the core levels in multinary materials. We show that the method can be used to construct the electronic band structure across a hetero junction. The results are then used as input into a microscopic theory allowing us to predict optical properties of the device. 

In \cref{sec:method} we describe the modified core-to-\gls{vbm} method applied for the calculation of the \gls{vbo}. We test the method for the example of the well-known GaAs/(Al,Ga)As interface in \cref{sec:algaas}. In \cref{sec:w}, we then determine the band alignments of the hetero junctions GaAs/Ga(As,Sb) and GaAs/(In,Ga)As. We use the resulting band structure as input into the microscopic calculation of the photoluminescence spectra of these systems which are currently under investigation for use in flexible type II semiconductor lasers~\cite{Fuchs2018}. The work is concluded in \cref{sec:discussion}.

\section{The modified core-to-VBM approach} 
\label{sec:method}

The \gls{vbo} of an interface $X/Y$ between two materials $X$ and $Y$ is defined as the \gls{vbm} energy difference of the layers, 
\begin{align}
    E_\mathrm{VBO} &= E^Y_v - E^X_v \, .
\end{align}
It is possible to extract $E^Y_v$ and $E^X_v$ from a single first-principles \gls{dft} calculation by calculating the position dependent \gls{vbm} (one value for each side $X$ and $Y$ of the interface), which can be done for instance by analyzing the \gls{ldos}~\cite{Bass1989}. However, such techniques introduce additional adjustable parameters and are relatively unprecise such that other approaches are usually preferred.

A more precise method is to compare the \glspl{vbm} $E_v^X$ and $E_v^Y$ of two separate calculations, one for each of the bulk materials $X$ and $Y$. Comparing the $E_v^X$ and $E_v^Y$ from separate bulk calculations has the advantage of yielding the true bulk-like band offset, as information about the interface is not included in the electronic structure calculations. In addition, the calculation of $E_v^X$ and $E_v^Y$ is much less involved, as the separate cells are smaller and finite size effects are less pronounced than in the interface-containing $X/Y$ cell.

\begin{figure}[t]
    \begin{center}
        \includegraphics[width=\linewidth]{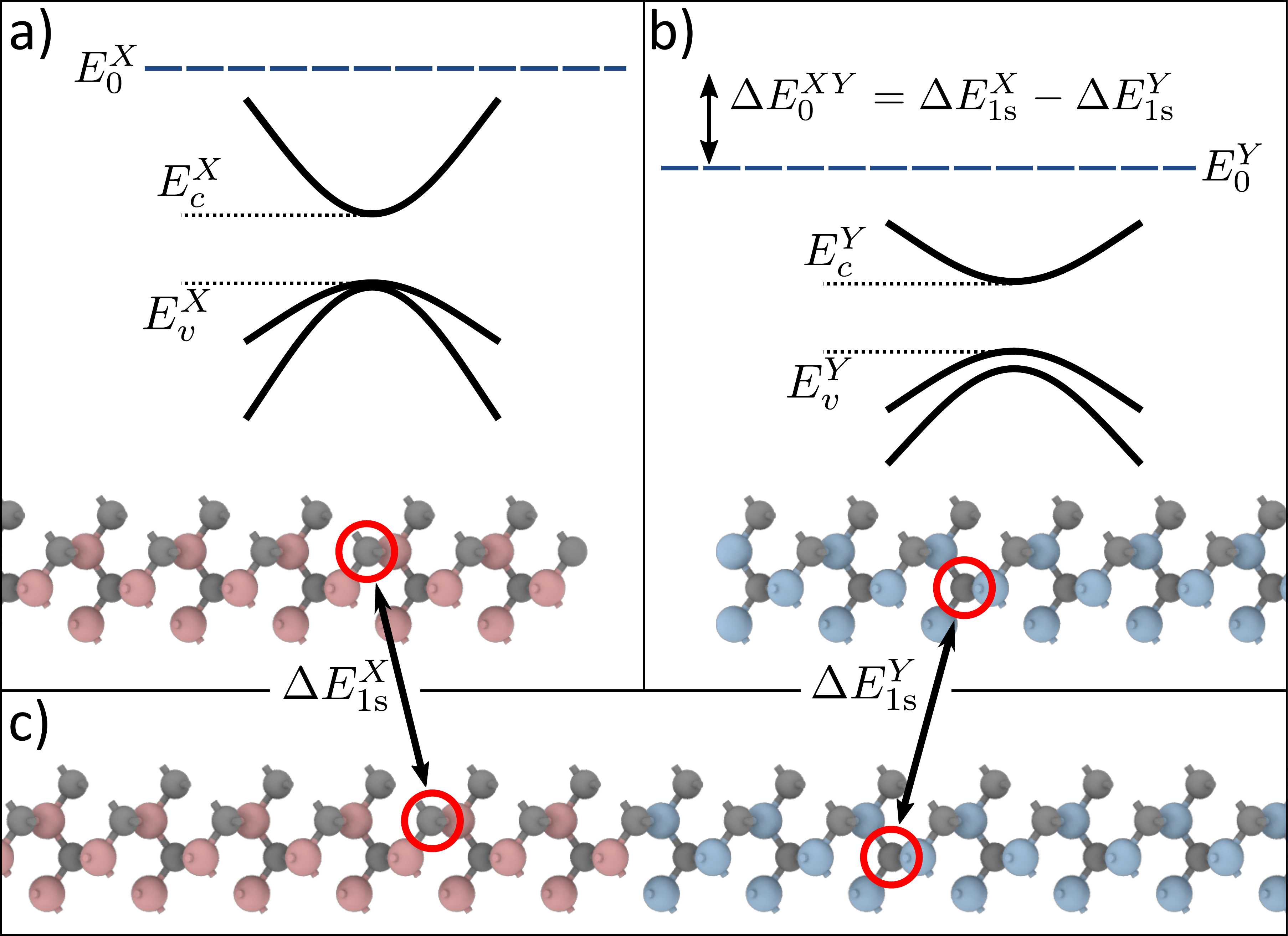}
        \caption{a) and b) Two bulk cells $X$ and $Y$ and schematic drawing of their band structures around the $\Gamma$-point, with the \glspl{vbm} $E_v^X$ and $E_v^Y$, and the conduction band minima $E_c^X$ and $E_c^Y$. $E_0^X$ and $E_0^Y$ are some reference energy level present in both cells (dashed blue lines). Because of the ill-defined average electrostatic potential in \gls{dft}, $X$ and $Y$ are misaligned by an energy offset of $\Delta E_0^{XY}$. In the modified core-level-to-\gls{vbm} approach, this offset is found by constructing a combined interface cell, shown in c). An ionic core state, here the 1s core level of the grey atoms, can then be compared between an atom in the bulk cell and its image in the interface cell (red circles), yielding $\Delta_\mathrm{1s}^X$ and $\Delta_\mathrm{1s}^Y$. The energy offset can then be calculated with $\Delta E_0^{XY} = \Delta_\mathrm{1s}^X - \Delta_\mathrm{1s}^Y$.}
        \label{fig:dft_method}
    \end{center}
\end{figure}

 However, since the average electrostatic potential in infinite solids is an ill-defined quantity~\cite{Balderschi1988a}, a common energy reference level in both systems is required to align their energy scales before $E_\mathrm{VBO}$ can be evaluated. \Cref{fig:dft_method} a) and b) show the two bulk cells $X$ and $Y$, their schematic band structure around the $\Gamma$-point, and some arbitrary reference level $E^X_0$ and $E_0^Y$ (blue, dashed lines). The energy scales of the two cells are misaligned by an offset $\Delta E_0^{XY}$. The problem of calculating the \gls{vbo} at the $X/Y$ interface therefore reduces to aligning the energy scales between the $X$ and $Y$ calculations, i.e., determining $\Delta E_0^{XY}$.

In the core-to-\gls{vbm} approach, ionic core levels $E_C^X$ and $E_C^Y$ provide the required reference energies~\cite{Wei1998}. These electronic states are sufficiently localized around the atomic cores not to be affected by ``distant'' structural features such as hetero interfaces. They do, however, depend on the chemical surrounding of their host ion, as well as the deformation potentials induced by lattice strain. For instance, the 1s core state of an As atom in GaAs is different to that of As atoms in AlAs. In multinary and strained materials, it becomes even more complicated as the core levels of any ion will depend on the exact chemical environment (distribution of atoms) and lattice parameters (strain and local static atomic displacements). It is therefore not possible to directly compare the core states of of the separate $X$ and $Y$ calculations. Fortunately, unlike the \glspl{vbm} $E_v^X$ and $E_v^Y$, core levels are not a bulk property but by nature very localized to their host ions, which makes it possible to calculate them in a single $X/Y$ interface calculation without loss of precision.

The $X/Y$ cell is constructed by stacking $X$ and $Y$ on top of each other, thereby simulating a perfectly abrupt interface between the two bulk cells. Consequently, each atom in the bulk cells $X$ and $Y$ has an image in the $X/Y$ cell with the exact same chemical surrounding (apart from ions very close to the interface) and similar static atomic displacements, i.e., deformation potentials. \Cref{fig:dft_method} c) shows the combined supercell and highlights example atoms in the individual bulk cells and their corresponding images in the $X/Y$ cell (red circles). 

The energy scales of $X$ and $Y$ are then aligned as follows: Some core level is calculated for \emph{all ions} in the bulk cells $X$ and $Y$ \emph{and} in the interface cell $X/Y$. A natural choice is the atomic 1s orbital, as it's radius is the smallest and it is conveniently symmetric. The corresponding energy levels are then $E_\mathrm{1s}^X$ and $E_\mathrm{1s}^Y$ for the ions in the $X$ and $Y$ cell, and their images $E_\mathrm{1s}^{X,X/Y}$, $E_\mathrm{1s}^{Y,X/Y}$ in the interface $X/Y$ cell. 

Because 
\begin{align}
    E_\mathrm{1s}^{X,X/Y} \overset{!}{=} E_\mathrm{1s}^{X} && \mathrm{and} &&
    E_\mathrm{1s}^{Y,X/Y} \overset{!}{=} E_\mathrm{1s}^{Y} \, ,
    \label{eq:vbo}
\end{align}
the energy offsets $\Delta E_\mathrm{1s}^X = E_\mathrm{1s}^{X,X/Y} - E_\mathrm{1s}^{X}$ and $\Delta E_\mathrm{1s}^Y = E_\mathrm{1s}^{Y,X/Y} - E_\mathrm{1s}^{Y}$ between each bulk cell and the $X/Y$ interface cell are found. The \gls{vbo} then is:

\begin{align}
    E_\mathrm{VBO} &= (E^Y_v - \Delta E_\mathrm{1s}^Y) - (E^X_v - \Delta E_\mathrm{1s}^X)\, .
    \label{eq:vbm}
\end{align}

It should be noted that in order to preserve the lattice strain and static atomic displacements in the $X/Y$ interface cell with respect to the two bulk calculations $X$ and $Y$, all three calculations need to be structurally relaxed before the calculation of the core states and the valence band maxima. In semiconductor hetero structures, the lattice parameters perpendicular to the growth direction are usually equal to the substrate lattice constant, and the layers are free to expand or shrink only in growth direction. This should be taken into account when applying the core-level-to-\gls{vbm} approach.

\section{The GaAs/(Al,Ga)As interface} 
\label{sec:algaas}

To test the accuracy of the modified core-to-\gls{vbm} method, we apply it to the GaAs/(Al$_\mathrm{x}$,Ga$_\mathrm{1-x}$)As hetero junction with varying composition x. Note, that throughout the paper, we study [001] growth direction and the corresponding interfaces. The (Al,Ga)As supercells are constructed using \gls{sqs}~\cite{VandeWalle2002} to approximate infinite bulk alloys. The 1s core levels of the ions are calculated using the initial state approximation~\cite{Kohler2004} and the \glspl{vbo} were found by determining the highest occupied energy state in the bulk cells, $E_v^\mathrm{GaAs}$ and $E_v^\mathrm{AlGaAs}$. Even though AlAs and GaAs have very similar lattice parameters, strain in the (Al,Ga)As cell was still taken into account by constraining the lattice constant parallel to the interface ($100$ and $010$ direction) to that of GaAs (\SI{0.566}{\nm}), and relaxing the cell in $001$ direction.

\begin{figure}[t]
    \begin{center}
        \includegraphics[width=\linewidth]{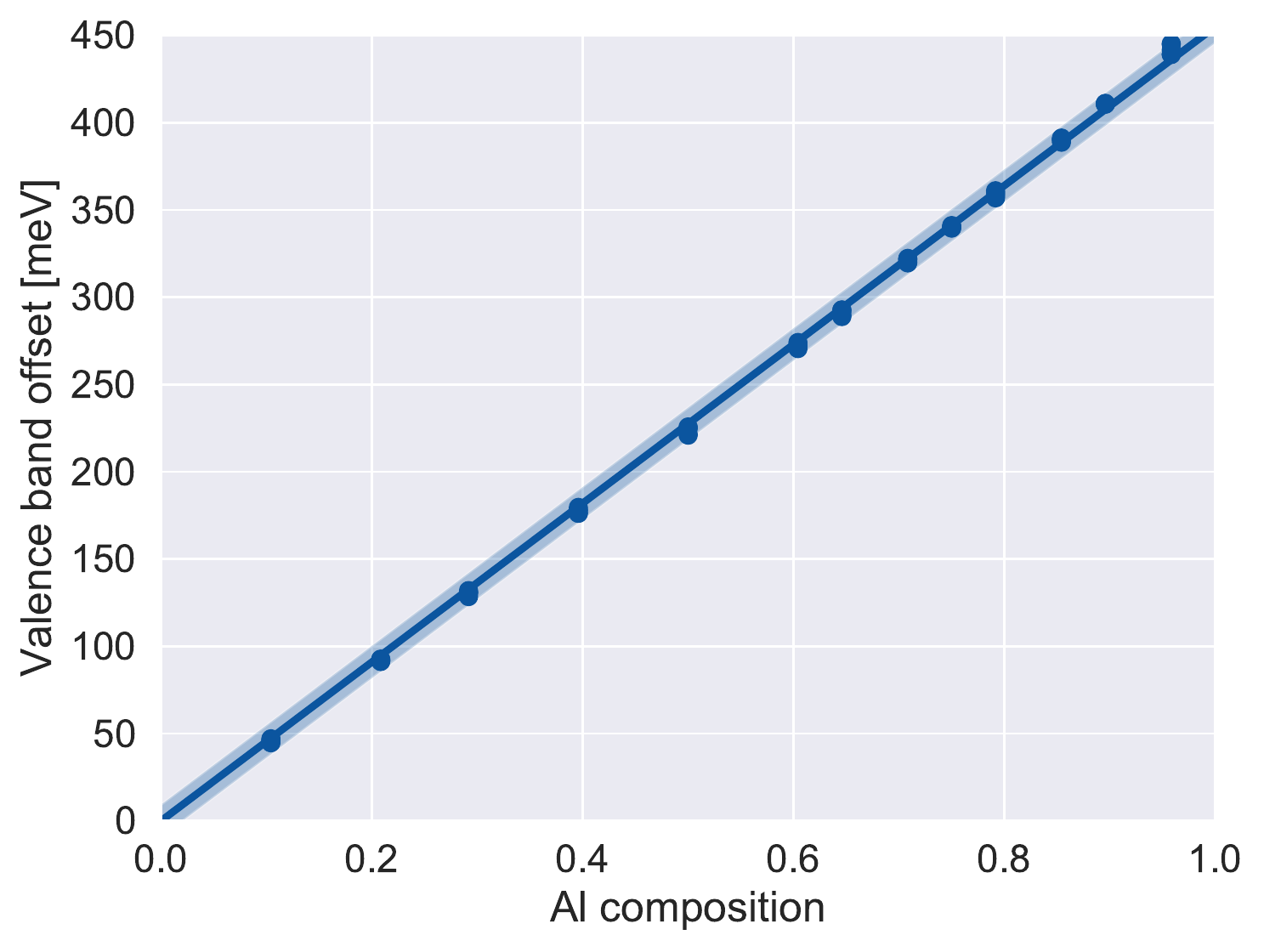}
        \caption{\gls{vbo} of the GaAs/(Al$_\mathrm{x}$,Ga$_\mathrm{1-x}$)As interface for different concentrations x of Al. The Al and Ga atoms in the upper cell were distributed using five realizations of \gls{sqs}, with the shaded area around the line fit representing the standard deviation of the \gls{vbo} with respect to the atomic distribution.}
        \label{fig:gaasalgaas_x}
    \end{center}
\end{figure}

\begin{figure}[t]
    \begin{center}
        \includegraphics[width=\linewidth]{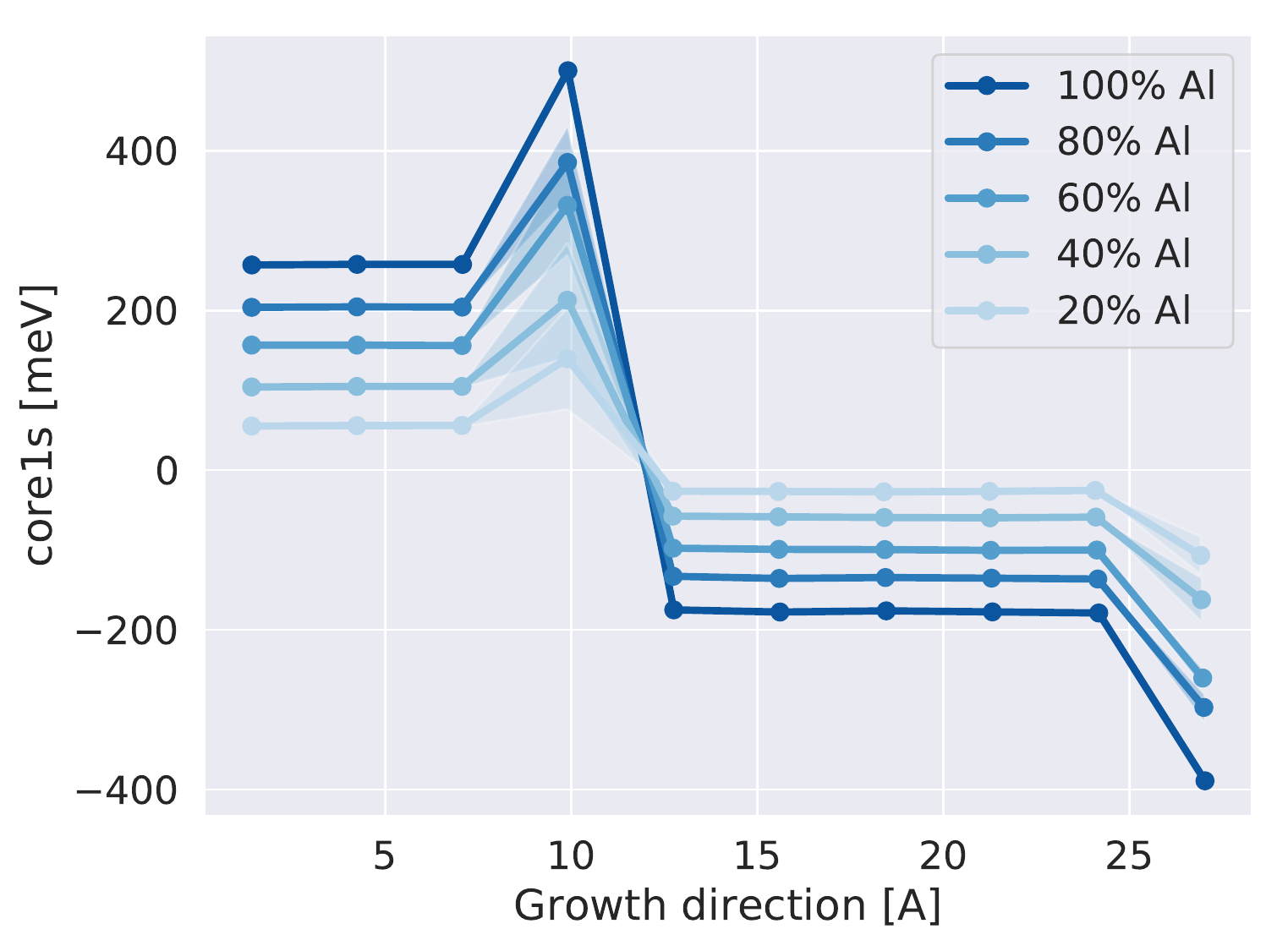}
        \caption{The difference between the 1s core states of As atoms in the GaAs/(Al$_\mathrm{x}$,Ga$_\mathrm{1-x}$)As cell with respect to the corresponding bulk reference states, as a function of position in growth direction.}
        \label{fig:core1s}
    \end{center}
\end{figure}

Let us briefly summarize the core-to-\gls{vbm} method described in \cref{sec:method} by the GaAs/(Al,Ga)As example.
\begin{enumerate}
    \itemsep0em 
    \item Construct the GaAs and (Al,Ga)As bulk cells with the GaAs lattice constant. Random distribution of Al and Ga atoms in the (Al,Ga)As cell is done by constructing \glspl{sqs}~\cite{VandeWalle2009}.
    \item Relax the bulk cells in (001) growth direction to account for strain.
    \item Run electronic structure calculations of the bulk cells to determine $E^\mathrm{AlGaAs}_v$ and $E^\mathrm{GaAs}_v$, as well as the 1s core levels $E^\mathrm{AlGaAs}_\mathrm{1s}$ and $E^\mathrm{GaAs}_\mathrm{1s}$.
    \item Stack the \emph{relaxed} bulk cells vertically ([001] direction) to construct the GaAs/(Al,Ga)As interface cell. 
    \item Relax the interface cell. This is done to take into account corrections of the atomic structure at the interface.
    \item Run an electronic structure calculation of the interface cell to obtain the core levels of the combined GaAs/(Al,Ga)As cell. Apply \cref{eq:vbo,eq:vbm} to find the \gls{vbo}. 
\end{enumerate}

For the \gls{dft} calculations, we use the pseudo-potential based Vienna Ab initio Simulation Package~\cite{Kresse1996} with the PBEsol exchange-correlation functional.~\cite{Perdew2007} Spin-orbit coupling is not taken into account for the electronic structure calculations of GaAs and (Al,Ga)As, as it has negligible effect on the position of the \glspl{vbm} in these materials~\cite{Komsa2008}. The size of the GaAs cell is $2\times 2\times 2$ cubic unit cells and we use a $\Gamma$-centered Monkhorst-Pack~\cite{Monkhorst1976} k-point grid of $5\times 5\times 5$. The (Al,Ga)As cell size is $2\times 2\times 3$ unit cells big (the long dimension is the growth direction) and a k-point grid of $5\times 5\times 3$ is used. For the calculation of the GaAs/(Al,Ga)As interface cell a $5\times 5\times 2$ k-point grid is used. The kinetic energy cut-off is \SI{368}{\electronvolt} in all calculations.

First, we study the \gls{vbo} between GaAs and (Al$_\mathrm{x}$,Ga$_\mathrm{1-x}$)As depending on Al concentration x, which was varied between $\SI{0}{\percent}$ and $\SI{100}{\percent}$. For each composition, five \gls{sqs} configurations of atom placement on the group III sublattice (Al and Ga) were calculated to estimate the impact different lattice realizations. The results are shown in \cref{fig:gaasalgaas_x}. We obtain a linear dependence of
\begin{align}
    E^\mathrm{GaAs/(Al,Ga)As}_\mathrm{VBO}(x_\mathrm{Al}) = x_\mathrm{Al}\times(\SI{455(10)}{\milli\electronvolt})\, .
\end{align}
Our result aligns well with previous literature values, all of which are in the range \SIrange{400}{550}{\milli\electronvolt} for the GaAs/AlAs interface~\cite{Vurgaftman2001,Wei1998,Wang1985}. A linear dependence of the offset on Al concentration has also been found previously~\cite{Wang1985}. The uncertainty related to the distribution of Al and Ga atoms in the (Al,Ga)As cell is \SI{12}{\milli\electronvolt}.

The position dependence of the 1s core states of As atoms throughout the supercell is shown in \cref{fig:core1s}. For each As (group III) layer in the hetero structure, the difference between the 1s core levels of the interface cell and of the bulk reference cell is shown. The energies are averaged over all As atoms in the layer. We can make two observations from the data. First, the core levels clearly converge to their bulk values within not more than two atomic layers distance from the interface. This shows that the interface has negligible effect on the core levels and that the core-to-\gls{vbm} method described in \cref{sec:method} requires not very extended cells in growth direction. Second, the relative core levels are constant in the respective parts of the cell, which indicates that they are not influenced by long-range electric fields, such as a macroscopic polarizations.\cite{Bernardini1998} 

Our results for the GaAs/(Al,Ga)As model system are encouraging and suggest that the method can readily be applied to other semiconductor hetero junctions. In the next section, we study the GaAs/Ga(As,Sb) and GaAs/(Ga,In)As interfaces, which are highly relevant for the construction of flexible type-II semiconductor lasers.

\section{GaAs/Ga(As,Sb)/(Ga,In)As MQW structures}
\label{sec:w}

The core-to-\gls{vbm} method is applied to the GaAs/(In,Ga)As and GaAs/Ga(As,Sb) hetero interfaces. These materials are currently investigated for use in \SI{1.3}{\micro\meter} type II semiconductor lasers.~\cite{Fuchs2018} The results are used to calculate \gls{pl} spectra for an (In,Ga)As/\allowbreak GaAs/\allowbreak Ga(As,Sb) \gls{mqw} structure, and are compared with experimental data from Ref.~\onlinecite{Gies2016}.

\subsection{Band offsets}

The details of the calculations are the same as for the GaAs/(Al,Ga)As interface: Realizations of atomic placements are generated using \gls{sqs}, the GaAs cell size is $2\times 2\times 2$ cubic unit cells and the (In,Ga)As and Ga(As,Sb) cells are $2\times 2\times 3$ unit cells big, with similar k-point grids as described in \cref{sec:algaas}. In all electronic structure calculations spin-orbit coupling is taken into account, as it has significant effect in In and Sb containing materials.

\begin{figure}[tb]
    \includegraphics[width=\linewidth]{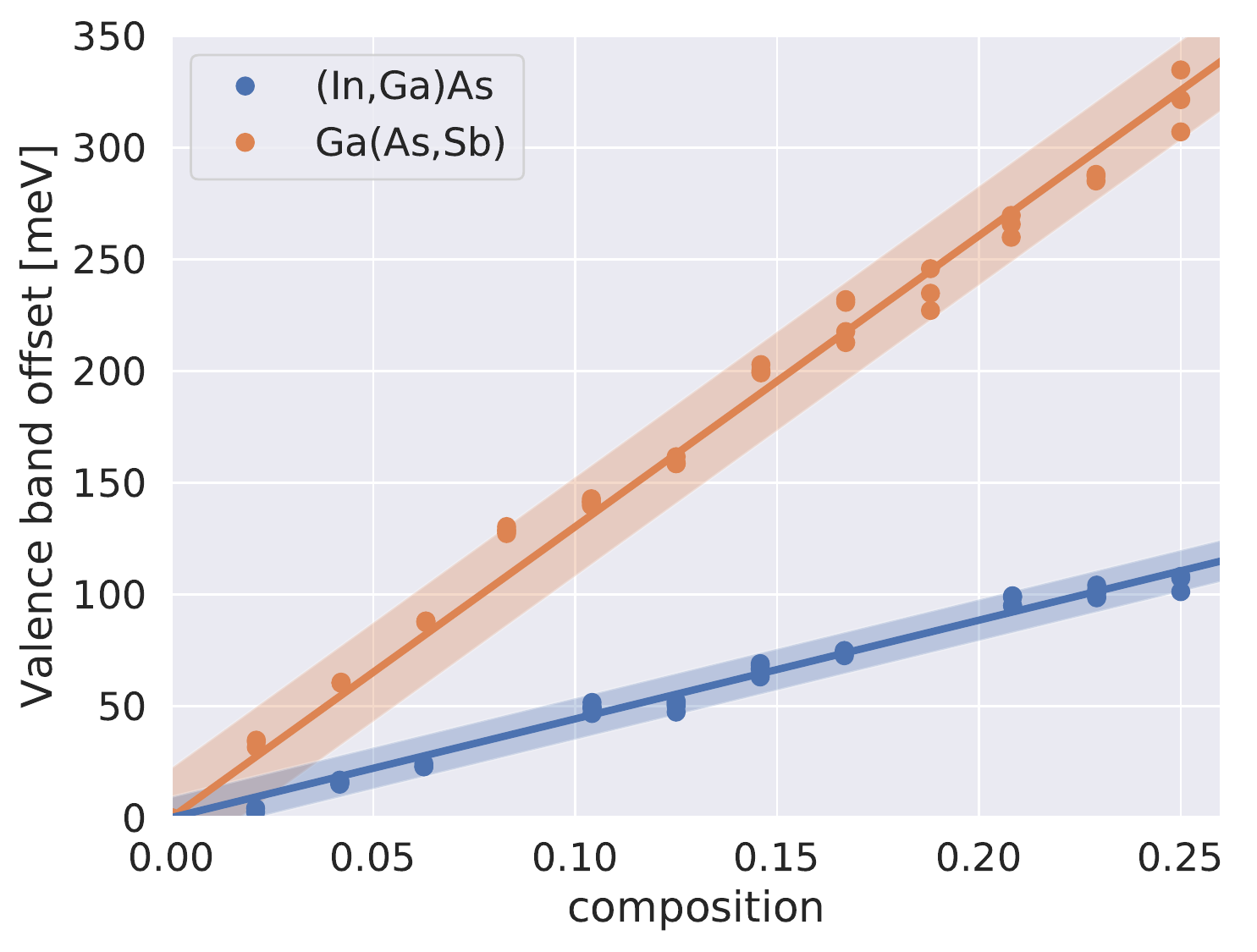}
    \caption{\glspl{vbo} of GaAs/(In,Ga)As and GaAs/Ga(As,Sb) versus concentration of In or Sb atoms in the ternary layer. Linear least squares fits are shown as straight lines, estimated deviations due to different placement of atoms in the (In,Ga)As and Ga(As,Sb) layers are shown as shaded areas.}
    \label{fig:InFit}
\end{figure}
 
For both interfaces, we study the dependence of the \gls{vbo} on concentration of the dilute constituents (In and Sb). The results are shown in \cref{fig:InFit}. Clearly, the \glspl{vbo} depend not only on the concentration of In or Sb, but vary also with atomic placement. In the studied composition range of \SIrange{0}{25}{\percent}, which covers the materials relevant for applications, the dependence of the offsets on concentration is linear. The data yields the following results for the composition-dependent \glspl{vbo}:
\begin{align}
    E^\mathrm{GaAs/(In,Ga)As}_\mathrm{VBO}(x_\mathrm{In}) &= x_\mathrm{In}\times(\SI{-440(10)}{\milli\electronvolt}) \nonumber \\
    E^\mathrm{GaAs/Ga(As,Sb)}_\mathrm{VBO}(x_\mathrm{Sb}) &= x_\mathrm{Sb}\times(\SI{-1300(20)}{\milli\electronvolt})\, .
    \label{eq:wres}
\end{align}

With the results from \cref{eq:wres} the (In,Ga)As/\allowbreak GaAs/\allowbreak Ga(As,Sb) \gls{mqw} structure can be constructed. The modeled structure is shown schematically in the inset of \cref{fig:PL}. \Cref{tab:mqw} lists the parameters for the different quantum wells, with compositions and thicknesses taken (within experimental uncertainty) from the study Ref.~\onlinecite{Gies2016}. 
The band gaps in \Cref{tab:mqw} are calculated using values from Ref.~\onlinecite{Vurgaftman2001}. For the calculation of the \glspl{vbo} the core-to-\gls{vbm} method, detailed above, is used. 

\begin{table}
    \begin{tabularx}{\linewidth}{ XXXXX }
        \toprule
                  & Thickness & Band gap & Conc. & \gls{vbo} \\ 
        \midrule
        (In,Ga)As & \SI{5.70}{\nano\meter} & \SI{1207}{\milli\electronvolt} & \SI{20.3}{\percent} In & \SI{89.3}{\milli\electronvolt} \\
        Ga(As,Sb) & \SI{5.5}{\nano\meter} & \SI{1103}{\milli\electronvolt} & \SI{23.7}{\percent} Sb & \SI{308}{\milli\electronvolt} \\
        GaAs      & \SI{4.80}{\nano\meter} & \SI{1447}{\milli\electronvolt} &                        & \\
        \bottomrule

        \end{tabularx}
    \caption{Parameters of the modeled \gls{mqw} structure. 
    \glspl{vbo} relative to GaAs are calculated from \cref{eq:wres}. Band gap values are for the band gaps in the strained MQW structure.} 
    \label{tab:mqw}
\end{table}


\subsection{Photoluminescence}

\begin{figure}[tb]
    \includegraphics[width=\linewidth]{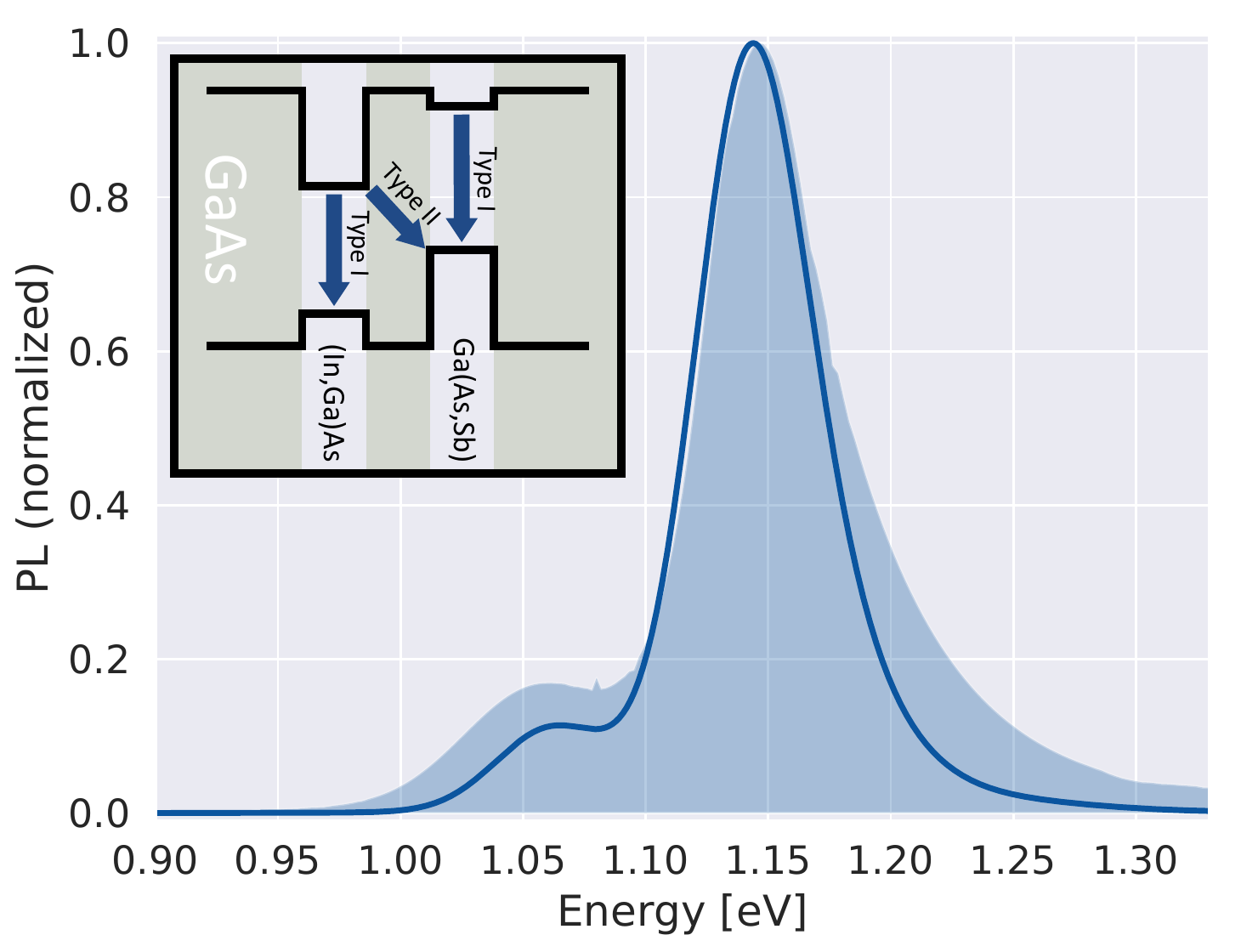}
    \caption{Normalized experimental (shaded area, from Ref.~\onlinecite{Gies2016}) and calculated (line) \gls{pl} spectra of the structure outlined in the inset. The type I and type II transitions of and between the two quantum wells and are clearly visible.}
    \label{fig:PL}
\end{figure}

Experimentally measured \gls{pl} spectra for GaAs/Ga(As,Sb)/(Ga,In)As MQW structures have been presented in Ref. \onlinecite{Gies2016}. A typical example is reproduced as the shaded area in Fig. \cref{fig:PL}. In the figure, we clearly see the type I transition of the Ga(As,Sb) quantum well around \SI{1.146}{\electronvolt} and the  
type II transition between the (In,Ga)As and Ga(As,Sb) quantum wells appears at \SI{1.056}{\electronvolt}, respectively. For more details on the transition assignment, the sample used, and the measurement techniques, we refer the interested reader to Ref. \onlinecite{Gies2016}.  

In our microscopic calculations of the \gls{pl} spectrum, we solve the semiconductor luminescence equations~\cite{Kira1999189}. Here, we assume equilibrium Fermi-Dirac distributions for the \SI{0.001e12}{\per\centi\meter\squared} excited electron-hole pairs. The Coulomb matrix elements, the dipole matrix elements, and the single particle energies are calculated using 8$\times$8 \textbf{k$\cdot$p} theory~\cite{Hader1997,Chow1994}. The electron-electron and electron-phonon scattering is taken into account at the level of the second-Born-Markov approximation~\cite{HaKoMo2003}. The spectrum is inhomogeneously broadened by convolution with a \SI{30}{\milli\electronvolt} FWHM Gaussian distribution in order to take structural disorder effects into account. 

The computed PL spectrum is shown as solid line in \cref{fig:PL}. The comparison with the experimental results yields good overall agreement. In particular, both the type I and the type II transitions are reproduced rather well, confirming that the computed band offsets are close to the experimentally realized values.

\section{Conclusion}
\label{sec:discussion}

In this paper we show how the core-to-\gls{vbm} method, modified to include anisotropic strain and \glspl{sad}, can be used to accurately predict \glspl{vbo} of III/V semiconductor hetero junctions. The method is first applied to the well-known GaAs/(Al,Ga)As interface and compares well with results from the literature. As expected, a linear dependence of the offset on Al concentration is found, with the GaAs/AlAs offset at \SI{455(10)}{\milli\electronvolt}.

In \cref{sec:w}, the approach is applied to the (In,Ga)As/\allowbreak GaAs/\allowbreak Ga(As,Sb) \gls{mqw}, which is currently investigated for flexible type II semiconductor lasers. Within the compositional uncertainty and for the region of interest of \SIrange{0}{25}{\percent} we find linear dependencies of the GaAs/(In,Ga)As and GaAs/Ga(As,Sb) \glspl{vbo} on In and Sb concentration, respectively. The calculated offsets are used to predict the \gls{pl} spectrum for an \gls{mqw} hetero structure, which is compared with experimental measurements from Ref.~\onlinecite{Gies2016}. The type II transition energy is well reproduced, indicating successful prediction of the offsets.

Our results show that the core-to-\gls{vbm} method is suitable to predict band alignments of strained semiconductor hetero junctions from first principles and can readily be applied to other material systems. In conjunction with solvers for optical properties, it is possible to calculate quantities such as \gls{pl} spectra without experimental input, allowing to theoretically design and tune functionalized semiconductor hetero structures prior to growth. By fitting to experimental measurements, such a combined approach can also be used to gain insight into structural or electronic properties of samples. 

\section{Acknowledgements}
Financial support is provided by the German Research Foundation (DFG) in the framework of the GRK 1782 and SFB 1083.


\begin{thebibliography}{24}%
\makeatletter
\providecommand \@ifxundefined [1]{%
 \@ifx{#1\undefined}
}%
\providecommand \@ifnum [1]{%
 \ifnum #1\expandafter \@firstoftwo
 \else \expandafter \@secondoftwo
 \fi
}%
\providecommand \@ifx [1]{%
 \ifx #1\expandafter \@firstoftwo
 \else \expandafter \@secondoftwo
 \fi
}%
\providecommand \natexlab [1]{#1}%
\providecommand \enquote  [1]{``#1''}%
\providecommand \bibnamefont  [1]{#1}%
\providecommand \bibfnamefont [1]{#1}%
\providecommand \citenamefont [1]{#1}%
\providecommand \href@noop [0]{\@secondoftwo}%
\providecommand \href [0]{\begingroup \@sanitize@url \@href}%
\providecommand \@href[1]{\@@startlink{#1}\@@href}%
\providecommand \@@href[1]{\endgroup#1\@@endlink}%
\providecommand \@sanitize@url [0]{\catcode `\\12\catcode `\$12\catcode
  `\&12\catcode `\#12\catcode `\^12\catcode `\_12\catcode `\%12\relax}%
\providecommand \@@startlink[1]{}%
\providecommand \@@endlink[0]{}%
\providecommand \url  [0]{\begingroup\@sanitize@url \@url }%
\providecommand \@url [1]{\endgroup\@href {#1}{\urlprefix }}%
\providecommand \urlprefix  [0]{URL }%
\providecommand \Eprint [0]{\href }%
\providecommand \doibase [0]{http://dx.doi.org/}%
\providecommand \selectlanguage [0]{\@gobble}%
\providecommand \bibinfo  [0]{\@secondoftwo}%
\providecommand \bibfield  [0]{\@secondoftwo}%
\providecommand \translation [1]{[#1]}%
\providecommand \BibitemOpen [0]{}%
\providecommand \bibitemStop [0]{}%
\providecommand \bibitemNoStop [0]{.\EOS\space}%
\providecommand \EOS [0]{\spacefactor3000\relax}%
\providecommand \BibitemShut  [1]{\csname bibitem#1\endcsname}%
\let\auto@bib@innerbib\@empty
\bibitem [{\citenamefont {Haug}\ and\ \citenamefont
  {Koch}(2009)}]{doi:10.1142/7184}%
  \BibitemOpen
  \bibfield  {author} {\bibinfo {author} {\bibfnamefont {H.}~\bibnamefont
  {Haug}}\ and\ \bibinfo {author} {\bibfnamefont {S.~W.}\ \bibnamefont
  {Koch}},\ }\href {\doibase 10.1142/7184} {\emph {\bibinfo {title} {{Quantum
  Theory of the Optical and Electronic Properties of Semiconductors}}}},\
  \bibinfo {edition} {5th}\ ed.\ (\bibinfo  {publisher} {World Scientific},\
  \bibinfo {year} {2009})\BibitemShut {NoStop}%
\bibitem [{\citenamefont {Winkler}\ and\ \citenamefont
  {R{\"{o}}ssler}(1993)}]{Winkler1993}%
  \BibitemOpen
  \bibfield  {author} {\bibinfo {author} {\bibfnamefont {R.}~\bibnamefont
  {Winkler}}\ and\ \bibinfo {author} {\bibfnamefont {U.}~\bibnamefont
  {R{\"{o}}ssler}},\ }\href {\doibase 10.1103/PhysRevB.48.8918} {\bibfield
  {journal} {\bibinfo  {journal} {Physical Review B}\ }\textbf {\bibinfo
  {volume} {48}},\ \bibinfo {pages} {8918} (\bibinfo {year}
  {1993})}\BibitemShut {NoStop}%
\bibitem [{\citenamefont {Franciosi}\ and\ \citenamefont {{Van De
  Walle}}(1996)}]{FRANCIOSI1996}%
  \BibitemOpen
  \bibfield  {author} {\bibinfo {author} {\bibfnamefont {A.}~\bibnamefont
  {Franciosi}}\ and\ \bibinfo {author} {\bibfnamefont {C.~G.}\ \bibnamefont
  {{Van De Walle}}},\ }\href {\doibase 10.1016/0167-5729(95)00008-9} {\bibfield
   {journal} {\bibinfo  {journal} {Surface Science Reports}\ }\textbf {\bibinfo
  {volume} {25}},\ \bibinfo {pages} {1} (\bibinfo {year} {1996})}\BibitemShut
  {NoStop}%
\bibitem [{\citenamefont {Tersoff}(1984)}]{Tersoff1984}%
  \BibitemOpen
  \bibfield  {author} {\bibinfo {author} {\bibfnamefont {J.}~\bibnamefont
  {Tersoff}},\ }\href {\doibase 10.1103/PhysRevB.30.4874} {\bibfield  {journal}
  {\bibinfo  {journal} {Physical Review B}\ }\textbf {\bibinfo {volume} {30}},\
  \bibinfo {pages} {4874} (\bibinfo {year} {1984})}\BibitemShut {NoStop}%
\bibitem [{\citenamefont {Wei}\ and\ \citenamefont {Zunger}(1998)}]{Wei1998}%
  \BibitemOpen
  \bibfield  {author} {\bibinfo {author} {\bibfnamefont {S.~H.}\ \bibnamefont
  {Wei}}\ and\ \bibinfo {author} {\bibfnamefont {A.}~\bibnamefont {Zunger}},\
  }\href {\doibase 10.1063/1.121249} {\bibfield  {journal} {\bibinfo  {journal}
  {Applied Physics Letters}\ }\textbf {\bibinfo {volume} {72}},\ \bibinfo
  {pages} {2011} (\bibinfo {year} {1998})}\BibitemShut {NoStop}%
\bibitem [{\citenamefont {{Van De Walle}}(2006)}]{VanDeWalle2006}%
  \BibitemOpen
  \bibfield  {author} {\bibinfo {author} {\bibfnamefont {C.~G.}\ \bibnamefont
  {{Van De Walle}}},\ }\href {\doibase 10.1016/j.physb.2005.12.004} {\bibfield
  {journal} {\bibinfo  {journal} {Physica B: Condensed Matter}\ }\textbf
  {\bibinfo {volume} {376-377}},\ \bibinfo {pages} {1} (\bibinfo {year}
  {2006})}\BibitemShut {NoStop}%
\bibitem [{\citenamefont {Komsa}\ \emph {et~al.}(2008)\citenamefont {Komsa},
  \citenamefont {Arola}, \citenamefont {Larkins},\ and\ \citenamefont
  {Rantala}}]{Komsa2008}%
  \BibitemOpen
  \bibfield  {author} {\bibinfo {author} {\bibfnamefont {H.-P.}\ \bibnamefont
  {Komsa}}, \bibinfo {author} {\bibfnamefont {E.}~\bibnamefont {Arola}},
  \bibinfo {author} {\bibfnamefont {E.}~\bibnamefont {Larkins}}, \ and\
  \bibinfo {author} {\bibfnamefont {T.~T.}\ \bibnamefont {Rantala}},\ }\href
  {\doibase 10.1088/0953-8984/20/31/315004} {\bibfield  {journal} {\bibinfo
  {journal} {Journal of Physics: Condensed Matter}\ }\textbf {\bibinfo {volume}
  {20}},\ \bibinfo {pages} {315004} (\bibinfo {year} {2008})}\BibitemShut
  {NoStop}%
\bibitem [{\citenamefont {Fuchs}\ \emph {et~al.}(2018)\citenamefont {Fuchs},
  \citenamefont {Br{\"{u}}ggemann}, \citenamefont {Weseloh}, \citenamefont
  {Berger}, \citenamefont {M{\"{o}}ller}, \citenamefont {Reinhard},
  \citenamefont {Hader}, \citenamefont {Moloney}, \citenamefont
  {B{\"{a}}umner}, \citenamefont {Koch},\ and\ \citenamefont
  {Stolz}}]{Fuchs2018}%
  \BibitemOpen
  \bibfield  {author} {\bibinfo {author} {\bibfnamefont {C.}~\bibnamefont
  {Fuchs}}, \bibinfo {author} {\bibfnamefont {A.}~\bibnamefont
  {Br{\"{u}}ggemann}}, \bibinfo {author} {\bibfnamefont {M.~J.}\ \bibnamefont
  {Weseloh}}, \bibinfo {author} {\bibfnamefont {C.}~\bibnamefont {Berger}},
  \bibinfo {author} {\bibfnamefont {C.}~\bibnamefont {M{\"{o}}ller}}, \bibinfo
  {author} {\bibfnamefont {S.}~\bibnamefont {Reinhard}}, \bibinfo {author}
  {\bibfnamefont {J.}~\bibnamefont {Hader}}, \bibinfo {author} {\bibfnamefont
  {J.~V.}\ \bibnamefont {Moloney}}, \bibinfo {author} {\bibfnamefont
  {A.}~\bibnamefont {B{\"{a}}umner}}, \bibinfo {author} {\bibfnamefont {S.~W.}\
  \bibnamefont {Koch}}, \ and\ \bibinfo {author} {\bibfnamefont
  {W.}~\bibnamefont {Stolz}},\ }\href {\doibase 10.1038/s41598-018-19189-1}
  {\bibfield  {journal} {\bibinfo  {journal} {Scientific Reports}\ }\textbf
  {\bibinfo {volume} {8}},\ \bibinfo {pages} {8} (\bibinfo {year}
  {2018})}\BibitemShut {NoStop}%
\bibitem [{\citenamefont {Bass}, \citenamefont {Oloumi},\ and\ \citenamefont
  {Matthai}(1989)}]{Bass1989}%
  \BibitemOpen
  \bibfield  {author} {\bibinfo {author} {\bibfnamefont {J.~M.}\ \bibnamefont
  {Bass}}, \bibinfo {author} {\bibfnamefont {M.}~\bibnamefont {Oloumi}}, \ and\
  \bibinfo {author} {\bibfnamefont {C.~C.}\ \bibnamefont {Matthai}},\ }\href
  {\doibase 10.1088/0953-8984/1/51/032} {\bibfield  {journal} {\bibinfo
  {journal} {Journal of Physics: Condensed Matter}\ }\textbf {\bibinfo {volume}
  {1}},\ \bibinfo {pages} {10625} (\bibinfo {year} {1989})}\BibitemShut
  {NoStop}%
\bibitem [{\citenamefont {Balderschi}, \citenamefont {Baroni},\ and\
  \citenamefont {Resta}(1988)}]{Balderschi1988a}%
  \BibitemOpen
  \bibfield  {author} {\bibinfo {author} {\bibfnamefont {A.}~\bibnamefont
  {Balderschi}}, \bibinfo {author} {\bibfnamefont {S.}~\bibnamefont {Baroni}},
  \ and\ \bibinfo {author} {\bibfnamefont {R.}~\bibnamefont {Resta}},\ }\href
  {\doibase 10.1103/PhysRevLett.61.734} {\bibfield  {journal} {\bibinfo
  {journal} {Physical Review Letters}\ }\textbf {\bibinfo {volume} {61}},\
  \bibinfo {pages} {734} (\bibinfo {year} {1988})}\BibitemShut {NoStop}%
\bibitem [{\citenamefont {{Van de Walle}}, \citenamefont {Asta},\ and\
  \citenamefont {Ceder}(2002)}]{VandeWalle2002}%
  \BibitemOpen
  \bibfield  {author} {\bibinfo {author} {\bibfnamefont {A.}~\bibnamefont {{Van
  de Walle}}}, \bibinfo {author} {\bibfnamefont {M.}~\bibnamefont {Asta}}, \
  and\ \bibinfo {author} {\bibfnamefont {G.}~\bibnamefont {Ceder}},\ }\href
  {\doibase 10.1016/S0364-5916(02)80006-2} {\bibfield  {journal} {\bibinfo
  {journal} {Calphad: Computer Coupling of Phase Diagrams and Thermochemistry}\
  }\textbf {\bibinfo {volume} {26}},\ \bibinfo {pages} {539} (\bibinfo {year}
  {2002})}\BibitemShut {NoStop}%
\bibitem [{\citenamefont {K{\"{o}}hler}\ and\ \citenamefont
  {Kresse}(2004)}]{Kohler2004}%
  \BibitemOpen
  \bibfield  {author} {\bibinfo {author} {\bibfnamefont {L.}~\bibnamefont
  {K{\"{o}}hler}}\ and\ \bibinfo {author} {\bibfnamefont {G.}~\bibnamefont
  {Kresse}},\ }\href {\doibase 10.1103/PhysRevB.70.165405} {\bibfield
  {journal} {\bibinfo  {journal} {Physical Review B}\ }\textbf {\bibinfo
  {volume} {70}},\ \bibinfo {pages} {165405} (\bibinfo {year}
  {2004})}\BibitemShut {NoStop}%
\bibitem [{\citenamefont {van~de Walle}(2009)}]{VandeWalle2009}%
  \BibitemOpen
  \bibfield  {author} {\bibinfo {author} {\bibfnamefont {A.}~\bibnamefont
  {van~de Walle}},\ }\href {\doibase 10.1016/j.calphad.2008.12.005} {\bibfield
  {journal} {\bibinfo  {journal} {Calphad: Computer Coupling of Phase Diagrams
  and Thermochemistry}\ }\textbf {\bibinfo {volume} {33}},\ \bibinfo {pages}
  {266} (\bibinfo {year} {2009})}\BibitemShut {NoStop}%
\bibitem [{\citenamefont {Kresse}\ and\ \citenamefont
  {Furthm{\"{u}}ller}(1996)}]{Kresse1996}%
  \BibitemOpen
  \bibfield  {author} {\bibinfo {author} {\bibfnamefont {G.}~\bibnamefont
  {Kresse}}\ and\ \bibinfo {author} {\bibfnamefont {J.}~\bibnamefont
  {Furthm{\"{u}}ller}},\ }\href {\doibase 10.1103/PhysRevB.54.11169} {\bibfield
   {journal} {\bibinfo  {journal} {Physical Review B: Condensed Matter and
  Materials Physics}\ }\textbf {\bibinfo {volume} {54}},\ \bibinfo {pages}
  {11169} (\bibinfo {year} {1996})}\BibitemShut {NoStop}%
\bibitem [{\citenamefont {Perdew}\ \emph {et~al.}(2008)\citenamefont {Perdew},
  \citenamefont {Ruzsinszky}, \citenamefont {Csonka}, \citenamefont {Vydrov},
  \citenamefont {Scuseria}, \citenamefont {Constantin}, \citenamefont {Zhou},\
  and\ \citenamefont {Burke}}]{Perdew2007}%
  \BibitemOpen
  \bibfield  {author} {\bibinfo {author} {\bibfnamefont {J.~P.}\ \bibnamefont
  {Perdew}}, \bibinfo {author} {\bibfnamefont {A.}~\bibnamefont {Ruzsinszky}},
  \bibinfo {author} {\bibfnamefont {G.~I.}\ \bibnamefont {Csonka}}, \bibinfo
  {author} {\bibfnamefont {O.~A.}\ \bibnamefont {Vydrov}}, \bibinfo {author}
  {\bibfnamefont {G.~E.}\ \bibnamefont {Scuseria}}, \bibinfo {author}
  {\bibfnamefont {L.~A.}\ \bibnamefont {Constantin}}, \bibinfo {author}
  {\bibfnamefont {X.}~\bibnamefont {Zhou}}, \ and\ \bibinfo {author}
  {\bibfnamefont {K.}~\bibnamefont {Burke}},\ }\href {\doibase
  10.1103/PhysRevLett.100.136406} {\bibfield  {journal} {\bibinfo  {journal}
  {Phys. Rev. Lett.}\ }\textbf {\bibinfo {volume} {100}},\ \bibinfo {pages}
  {136406} (\bibinfo {year} {2008})}\BibitemShut {NoStop}%
\bibitem [{\citenamefont {Monkhorst}\ and\ \citenamefont
  {Pack}(1976)}]{Monkhorst1976}%
  \BibitemOpen
  \bibfield  {author} {\bibinfo {author} {\bibfnamefont {H.~J.}\ \bibnamefont
  {Monkhorst}}\ and\ \bibinfo {author} {\bibfnamefont {J.~D.}\ \bibnamefont
  {Pack}},\ }\href {\doibase 10.1103/PhysRevB.13.5188} {\bibfield  {journal}
  {\bibinfo  {journal} {Physical Review B}\ }\textbf {\bibinfo {volume} {13}},\
  \bibinfo {pages} {5188} (\bibinfo {year} {1976})}\BibitemShut {NoStop}%
\bibitem [{\citenamefont {Vurgaftman}, \citenamefont {Meyer},\ and\
  \citenamefont {Ram-Mohan}(2001)}]{Vurgaftman2001}%
  \BibitemOpen
  \bibfield  {author} {\bibinfo {author} {\bibfnamefont {I.}~\bibnamefont
  {Vurgaftman}}, \bibinfo {author} {\bibfnamefont {J.~R.}\ \bibnamefont
  {Meyer}}, \ and\ \bibinfo {author} {\bibfnamefont {L.~R.}\ \bibnamefont
  {Ram-Mohan}},\ }\href {\doibase 10.1063/1.1368156} {\bibfield  {journal}
  {\bibinfo  {journal} {Journal of Applied Physics}\ }\textbf {\bibinfo
  {volume} {89}},\ \bibinfo {pages} {5815} (\bibinfo {year}
  {2001})}\BibitemShut {NoStop}%
\bibitem [{\citenamefont {Wang}\ and\ \citenamefont {Stern}(1985)}]{Wang1985}%
  \BibitemOpen
  \bibfield  {author} {\bibinfo {author} {\bibfnamefont {W.~I.}\ \bibnamefont
  {Wang}}\ and\ \bibinfo {author} {\bibfnamefont {F.}~\bibnamefont {Stern}},\
  }\href {\doibase 10.1116/1.583012} {\bibfield  {journal} {\bibinfo  {journal}
  {Journal of Vacuum Science {\&} Technology B}\ }\textbf {\bibinfo {volume}
  {3}},\ \bibinfo {pages} {1280} (\bibinfo {year} {1985})}\BibitemShut
  {NoStop}%
\bibitem [{\citenamefont {Bernardini}\ and\ \citenamefont
  {Fiorentini}(1998)}]{Bernardini1998}%
  \BibitemOpen
  \bibfield  {author} {\bibinfo {author} {\bibfnamefont {F.}~\bibnamefont
  {Bernardini}}\ and\ \bibinfo {author} {\bibfnamefont {V.}~\bibnamefont
  {Fiorentini}},\ }\href {\doibase 10.1103/PhysRevB.57.R9427} {\bibfield
  {journal} {\bibinfo  {journal} {Physical Review B - Condensed Matter and
  Materials Physics}\ }\textbf {\bibinfo {volume} {57}},\ \bibinfo {pages}
  {R9427} (\bibinfo {year} {1998})}\BibitemShut {NoStop}%
\bibitem [{\citenamefont {Gies}\ \emph {et~al.}(2016)\citenamefont {Gies},
  \citenamefont {Weseloh}, \citenamefont {Fuchs}, \citenamefont {Stolz},
  \citenamefont {Hader}, \citenamefont {Moloney}, \citenamefont {Koch},\ and\
  \citenamefont {Heimbrodt}}]{Gies2016}%
  \BibitemOpen
  \bibfield  {author} {\bibinfo {author} {\bibfnamefont {S.}~\bibnamefont
  {Gies}}, \bibinfo {author} {\bibfnamefont {M.~J.}\ \bibnamefont {Weseloh}},
  \bibinfo {author} {\bibfnamefont {C.}~\bibnamefont {Fuchs}}, \bibinfo
  {author} {\bibfnamefont {W.}~\bibnamefont {Stolz}}, \bibinfo {author}
  {\bibfnamefont {J.}~\bibnamefont {Hader}}, \bibinfo {author} {\bibfnamefont
  {J.~V.}\ \bibnamefont {Moloney}}, \bibinfo {author} {\bibfnamefont {S.~W.}\
  \bibnamefont {Koch}}, \ and\ \bibinfo {author} {\bibfnamefont
  {W.}~\bibnamefont {Heimbrodt}},\ }\href {\doibase 10.1063/1.4968541}
  {\bibfield  {journal} {\bibinfo  {journal} {Journal of Applied Physics}\
  }\textbf {\bibinfo {volume} {120}},\ \bibinfo {pages} {204303} (\bibinfo
  {year} {2016})}\BibitemShut {NoStop}%
\bibitem [{\citenamefont {Kira}\ \emph {et~al.}(1999)\citenamefont {Kira},
  \citenamefont {Jahnke}, \citenamefont {Hoyer},\ and\ \citenamefont
  {Koch}}]{Kira1999189}%
  \BibitemOpen
  \bibfield  {author} {\bibinfo {author} {\bibfnamefont {M.}~\bibnamefont
  {Kira}}, \bibinfo {author} {\bibfnamefont {F.}~\bibnamefont {Jahnke}},
  \bibinfo {author} {\bibfnamefont {W.}~\bibnamefont {Hoyer}}, \ and\ \bibinfo
  {author} {\bibfnamefont {S.~W.}\ \bibnamefont {Koch}},\ }\href {\doibase
  http://dx.doi.org/10.1016/S0079-6727(99)00008-7} {\bibfield  {journal}
  {\bibinfo  {journal} {Progress in Quantum Electronics}\ }\textbf {\bibinfo
  {volume} {23}},\ \bibinfo {pages} {189} (\bibinfo {year} {1999})}\BibitemShut
  {NoStop}%
\bibitem [{\citenamefont {Hader}, \citenamefont {Linder},\ and\ \citenamefont
  {D{\"{o}}hler}(1997)}]{Hader1997}%
  \BibitemOpen
  \bibfield  {author} {\bibinfo {author} {\bibfnamefont {J.}~\bibnamefont
  {Hader}}, \bibinfo {author} {\bibfnamefont {N.}~\bibnamefont {Linder}}, \
  and\ \bibinfo {author} {\bibfnamefont {G.~H.}\ \bibnamefont {D{\"{o}}hler}},\
  }\href {\doibase 10.1103/PhysRevB.55.6960} {\bibfield  {journal} {\bibinfo
  {journal} {Physical Review B}\ }\textbf {\bibinfo {volume} {55}},\ \bibinfo
  {pages} {6960} (\bibinfo {year} {1997})}\BibitemShut {NoStop}%
\bibitem [{\citenamefont {Chow}, \citenamefont {Koch},\ and\ \citenamefont
  {Sargent}(1994)}]{Chow1994}%
  \BibitemOpen
  \bibfield  {author} {\bibinfo {author} {\bibfnamefont {W.~W.}\ \bibnamefont
  {Chow}}, \bibinfo {author} {\bibfnamefont {S.~W.}\ \bibnamefont {Koch}}, \
  and\ \bibinfo {author} {\bibfnamefont {M.}~\bibnamefont {Sargent}},\ }\href
  {\doibase 10.1007/978-3-642-61225-1} {\emph {\bibinfo {title}
  {{Semiconductor-Laser Physics}}}}\ (\bibinfo  {publisher} {Springer Berlin
  Heidelberg},\ \bibinfo {year} {1994})\BibitemShut {NoStop}%
\bibitem [{\citenamefont {Hader}, \citenamefont {Koch},\ and\ \citenamefont
  {Moloney}(2003)}]{HaKoMo2003}%
  \BibitemOpen
  \bibfield  {author} {\bibinfo {author} {\bibfnamefont {J.}~\bibnamefont
  {Hader}}, \bibinfo {author} {\bibfnamefont {S.}~\bibnamefont {Koch}}, \ and\
  \bibinfo {author} {\bibfnamefont {J.~V.}\ \bibnamefont {Moloney}},\ }\href
  {\doibase 10.1016/S0038-1101(02)00405-7} {\bibfield  {journal} {\bibinfo
  {journal} {Solid-State Electron.}\ }\textbf {\bibinfo {volume} {47}},\
  \bibinfo {pages} {513} (\bibinfo {year} {2003})}\BibitemShut {NoStop}%
\end{thebibliography}
\end{document}